\documentstyle[12pt]{article}

\newcommand{\vt}[1]{\mbox{\ignorespaces\boldmath$#1$}}
\title{Calculation of the Electroelastic Green's Function
of the Hexagonal Infinite Medium}
\author{Thomas Michelitsch \\  Institut f\"ur Werkstoffwissenschaft \\
    TU Dresden\\
     D-01069 Dresden\\ Germany }
     \begin{document}
\begin{center}
{\it Thomas Michelitsch\footnote{\noindent Present address: \\
Sorbonne Universit\'es \\
Universit\'e Pierre et Marie Curie (Paris VI) \\
Institut Jean le Rond d'Alembert, CNRS UMR 7190 \\
Case 162 (4 Place Jussieu, Tour 55-65) \\
75252 Paris cedex 05 \\
FRANCE \\
Tél. +33 144 274928 \\
Fax. +33 144 275259 \\
Email: michel@lmm.jussieu.fr \\
http://bit.ly/champs-fractelysees },
Calculation of the Electroelastic Green's Function
of the Hexagonal Infinite Medium \\
Zeitschrift f\"ur Physik B Condensed Matter 10/1997; 104(3):497-503.\\ DOI: 10.1007/s002570050481 }
\end{center}
\maketitle
\begin{abstract}
The electroelastic $4 \times 4$ Green's function of a piezoelectric hexagonal (trans\-versely iso\-tropic) infinitely
extended medium is calculated explicitely in closed {\it compact} form
(eqs. (\ref{Greenform}) ff. and (\ref{Greentensor}) ff., respectively)
by using residue calculation. The results
can also be derived from Fredholm's method \cite{Fredholm}.
In the case of {\it vanishing} piezoelectric coupling the derived Green's function
coincides with two well known results: Kr\"oner's
expressions for the elastic Green's function tensor \cite{Kroener4} is reproduced
and the electric part then coincides with the electric potential
(solution of Poisson equation) which is caused by a unit point charge. \\
The obtained electroelastic Green's function is useful
for the calculation of the electroelastic Eshelby tensor \cite{KuhnMichel}.\\ \\
\footnotesize{\itemsep=-.1cm\frenchspacing
}\\ \\
Suggested PACS index: 77.22.-d
\end{abstract}

\section{Introduction}
In the solution of many problems in physics Green's functions
or fundamental solutions play an essential role.
The elastic Green's function tensor has been derived for isotropic
media by Sir Thompson \cite{Kelvin} and implicitely for anisotropic media
by Fredholm \cite{Fredholm}. \\
Especially the elastic Green's function of the infinite hexagonal medium
has been calculated explicitely by Lifshitz and Rosenzweig \cite{Lifshitz}
and by Kr\"oner \cite{Kroener4}. \\ \\
A first treatment of the electroelastic Green's function has been given
by Deeg \cite{Deeg}. Further contributions have been presented
by Dunn
\cite{Dunn}, Wang \cite{Wang} and Huang and Yu \cite{HuYu}
in the framework of inclusion problems.
But no compact explicit form for the electroelastic Green's function
has been given there. Thus, according to the author's knowledge,
a {\it compact} closed form representation
of the electroelastic Green's function does not exist.
But, due to the widespread interest in piezoelectric materials
a compact explicit representation
is highly desirable. \\
The residue method which is applied here is a useful tool to obtain
Green's functions in media with hexagonal symmetry as shown by Michelitsch
\cite{Michelitsch} and
Michelitsch and Wunderlin \cite{MiWuA} for a treatment of the incompatibility
problem. \\
In the following we calculate using this residue method
the electroelastic $4 \times 4$ Green's function for the hexagonal infinite piezoelectric
medium. We shall obtain this Green's function analytically in closed form
(eqs. (\ref{Greenform}) ff. and (\ref{Greentensor}) ff., respectively).
To obtain a compact formulation the use of a convenient symmetric
tensor basis is essential \cite{Levin,LevKre}. The basis tensors
used here are very useful for the solution of several problems in
hexagonal media \cite{Michelitsch,MiWu,MiWuD}. \\
The result for the electroelastic Green's function
presented here may be useful for the treatment of many problems, e.g.
the inclusion problem in piezoelectric hexagonal (transversely isotropic) media.

\section{Basic Equations}
We start from the field equations for the stress tensor ${\vt \sigma}$ and the dielectric
displacements ${\vt D}$.
The equilibrium conditions for the stresses are:

\begin{equation}
\label{eq}
\partial_j\sigma_{ij}=-K_i \,,\,\,\, \sigma_{ij} =\sigma_{ji}
\end{equation}

$\partial_j$ indicate the spacial derivatives and
${\vt K}$ the density of body forces, respectively. A further field equation describes the
conservation of free electric charges:

\begin{equation}
\label{lo}
\partial_lD_l =\rho_{e}
\end{equation}

$\rho_{e}$ represents the density of free electric charges.
The constitutive equations (material law) of the piezoelectric medium
which connect the electric field ${\vt E}$ and
the elastic deformation ${\vt \epsilon}$ with the
dielectric displacement ${\vt D}$ and the stress ${\vt \sigma}$ are given by:

\begin{equation}
\label{constlaw}
\begin{array}{lcc}
\sigma _{ij}&=&{\cal C}_{ijkl}\epsilon _{kl}-e_{kij}E_k \\
D_i&=&e_{ikl}\epsilon _{kl}+\eta _{ik}E_k \\
\end{array}
\end{equation}

${\cal C}_{ijkl}$, $e_{kij}$ and $\eta_{ij}$ denote the elastic moduli, the
piezoelectric moduli and the dielectric moduli, respectively.
They have the symmetry properties
${\cal C}_{ijkl}={\cal C}_{klij}={\cal C}_{jikl}={\cal C}_{ijlk}$, $e_{ikl}=e_{ilk}$ and
$\eta_{ij}=\eta_{ji}$.
Introducing the electric potential $\Phi$ and the elastic displacement field
${\vt u}$ the ansatz for the strain ${\vt \epsilon}$ and the electric
field ${\vt E}$ is given by:

\begin{equation}
\label{fieldans}
\begin{array}{lcc}
\epsilon_{ij}&=&\frac{1}{2}\left(\partial_iu_j+\partial_ju_i\right)\\
E_i&=&-\partial_i\Phi \\
\end{array}
\end{equation}

Putting ansatz (\ref{fieldans}) using (\ref{constlaw}) into
the field equations (\ref{eq}) and (\ref{lo}) we obtain
a $4 \times 4$ differential equation of degree two for the field
${\vt {\cal U}}=({\vt u},\Phi)$ of the form:

\begin{equation}
\label{tensdef}
{\vt {\cal T}}\left(\nabla\right){\vt {\cal U}} + {\vt {\cal F}}=0
\end{equation}

$\nabla$ indicates the gradient operator.
Here we have introduced the generalized force density
${\vt {\cal F}}=({\vt K},-\rho_{e})$. The symmetric $4 \times 4$
second order differential operator
${\vt {\cal T}}\left(\nabla\right)$ can be
written in the form:

\begin{equation}
\label{tensop}
\begin{array}{lcrr}
{\vt {\cal T}}\left(\nabla\right) & = &
\left[ \begin{array}{cc}
        {\vt T}\left(\nabla\right) & {\vt t}\left(\nabla\right)  \\
        {\vt t}^T\left(\nabla\right) & {\vt \tau}\left(\nabla\right)
\end{array} \right]
\end{array}
\end{equation}

Here ${\vt T}\left(\nabla\right)$ is a $3 \times 3$ tensor operator
and represents the elastic part and is given by:

\begin{equation}
\label{elast}
T_{ij}\left(\nabla\right) = {\cal C}_{ipjq}\partial_p\partial_q
\end{equation}

${\vt t}\left(\nabla\right)$ is a ($3 \times 1$ tensor) vector operator given by

\begin{equation}
\label{piez}
t_i\left(\nabla\right) = e_{piq}\partial_p\partial_q
\end{equation}
and represents the piezoelectric coupling.

Finally the ($1 \times 1$ tensor) scalar operator

\begin{equation}
\label{diel}
\tau\left(\nabla\right) = -\eta_{pq}\partial_p\partial_q
\end{equation}
describes the dielectric part. \\
The vector field ${\vt {\cal U}}$ can then be represented by the
$4 \times 4$ electroelastic Green's function ${\vt {\cal G}}$
according to

\begin{equation}
\label{sol}
{\vt {\cal U}}\left({\vt r}\right)=
{\displaystyle \int{{\vt {\cal G}}\left({\vt r}-{\vt r}^{'}\right)
{\vt {\cal F}}\left({\vt r}^{'}\right){\rm d}^3{\vt r}^{'}}}
\end{equation}

${\vt r}$ denotes the space point.
The electroelastic Green's function ${\vt {\cal G}}$ then is defined according to

\begin{equation}
\label{Greendef}
{\vt {\cal T}}\left(\nabla\right){\vt {\cal G}}\left({\vt r}\right)+
\delta^3\left({\vt r}\right){\vt 1}=0
\end{equation}

$\delta^3\left({\vt r}\right)$ represents the three-dimensional
$\delta$-function and ${\vt 1}$ denotes the $4 \times 4$ unit matrix. \\ \\
The electroelastic Green's function ${\cal G}_{pq}$ ($p,q=1,2,3,4$) has the
following physical interpretation \cite{Dunn}: \\
${\cal G}_{mj}\left({\vt r}\right)$ ($m,j=1,2,3$)
is the elastic displacement at spacepoint ${\vt r}$ in the $m$-direction
caused by a unit point force
at spacepoint ${\vt r}^{'}=0$ in the $j$-direction; \\
${\cal G}_{m4}\left({\vt r}\right)$ ($m=1,2,3$) is the elastic displacement
at spacepoint ${\vt r}$
in $m$-direction caused by a unit point charge at spacepoint ${\vt r}^{'}=0$;\\
${\cal G}_{4j}\left({\vt r}\right)$ ($j=1,2,3$) is the electric potential at
spacepoint ${\vt r}$ caused by a unit point force at spacepoint
${\vt r}^{'}=0$ in the $j$-direction;\\
${\cal G}_{44}\left({\vt r}\right)$ is the electric potential at spacepoint
${\vt r}$ caused by a unit point charge at spacepoint ${\vt r}^{'}=0$. \\
\\
We note that our electroelastic Green's function defined in equation
(\ref{Greendef}) is symmetric, i. e.
${\cal G}_{pq}={\cal G}_{qp}$. This property follows from the symmetry
of the operator ${\cal T}_{rs}\left(\nabla\right)=
{\cal T}_{sr}\left(\nabla\right)$ from equation (\ref{tensop}).
For our further calculation we use the following convenient
representation for the Green's function \cite{Deeg,Levin,LevKre,Michelitsch,MiWuA}:

\begin{equation}
\label{repres}
{\vt {\cal G}}\left({\vt r}\right) = {\displaystyle \frac{1}{8\pi^2r}}
\int_0^{2\pi}{{\vt {\cal T}}^{-1}\left({\vt \xi}\left(\alpha\right)\right)
{\rm d}\alpha}
\end{equation}

This relation can be derived from (\ref{Greendef}) in straight forward
manner by using Fourier transformation \cite{Michelitsch,MiWuA}.
One gets ${\vt {\cal T}}\left({\vt \xi}\right)$ from equation (\ref{tensop})
by replacing $\partial_i$
by $\xi_i$ ($i=1,2,3$) in equations (\ref{elast}), (\ref{piez}) and (\ref{diel}).
The vector ${\vt \xi}$ is then given by \cite{Michelitsch,MiWuA}

\begin{equation}
\label{xi}
{\vt \xi}\left(\alpha\right)= {\vt e}_1\cos{\alpha}+{\vt e}_2\sin{\alpha}
\end{equation}

The vectors ${\vt e}_i$ form a useful orthonormal basis

\begin{equation}
\label{evec}
\begin{array}{lcrclcrclcr}
  {\vt e}_1 & = & \frac{\displaystyle 1}{\displaystyle \rho}
  \left( \begin{array}{c}
       -y \\ x \\ 0
\end{array}    \right) &,&
  {\vt e}_2 & = &\frac{\displaystyle 1}{\displaystyle \rho r}
  \left( \begin{array}{c}
     -zx \\ -zy \\ \rho^2
\end{array}    \right) &,&
  {\vt e}_3 & = & \frac{\displaystyle 1}{\displaystyle r}
  \left( \begin{array}{c}
             x \\ y \\ z
\end{array}    \right) ,\\
\end{array}
\end{equation}
where $\rho^2=x^2+y^2$, $r^2=\rho^2+z^2$ and ${\vt r}=r{\vt e}_3$.
The orientation of this coordinate system can be expressed by
${\vt e}_i = \frac{\displaystyle 1}{\displaystyle 2}
\epsilon_{ijk}{\vt e}_j\times{\vt e}_k$. $\epsilon_{ijk}$ denotes
the antisymmetric permutation tensor.

\section{Residue Calculation}
Our goal is to formulate the residue calculation ansatz
to obtain the Green's function
from equation (\ref{repres}).\\
First of all we introduce for convenience a useful orthonormal basis to
represent the $4 \times 4$ matrix ${\vt {\cal T}}\left({\vt \xi}\right)$
and later ${\vt {\cal T}}^{-1}\left({\vt \xi}\right)$:

\begin{equation}
\label{hexbasis}
\begin{array}{lcrclcrclcrclcr}
  {\vt e}_b & = & \frac{\displaystyle 1}{\displaystyle \xi_b}
  \left( \begin{array}{c}
       \xi_1 \\ \xi_2 \\ 0\\0
\end{array}    \right) &,&
  {\vt e}_{b\perp} & = &\frac{\displaystyle 1}{\displaystyle \xi_b}
  \left( \begin{array}{c}
     -\xi_2 \\ \xi_1 \\ 0 \\ 0
\end{array}    \right) &,&
  {\vt e}_c & = &
  \left( \begin{array}{c}
             0 \\ 0 \\ 1 \\0
\end{array}    \right) &,&
  {\vt e}_4 & = &
  \left( \begin{array}{c}
             0 \\ 0 \\ 0 \\1
\end{array}    \right) \\
\end{array}
\end{equation}

($\xi_{\rm b}=\sqrt{\xi_1^2+\xi_2^2}$, $\xi_{\rm c}=\xi_3$).
This basis represents the hexagonal symmetry.
${\vt e}_b$ and ${\vt e}_{b\perp}$ are parallel to the basal plane
and ${\vt e}_c$ represents the $c$-direction. ${\vt e}_4$ comes into play
because of the electric potential component $\Phi$.
We then obtain the following useful representation:

\begin{equation}
\label{userep}
\begin{array}{lll}
{\vt {\cal T}}\left({\vt \xi}\right) &=&
T_{{\rm b}\perp}{\vt e}_{b\perp}\otimes{\vt e}_{b\perp}
+T_{\rm b}{\vt e}_{b}\otimes{\vt e}_{b}
+T_{\rm{bc}}\left({\vt e}_{b}\otimes{\vt e}_{c}+
{\vt e}_{c}\otimes{\vt e}_{b}\right)
+T_{\rm c}{\vt e}_{c}\otimes{\vt e}_{c} \nonumber \\ &&
+t_{\rm{b4}}\left({\vt e}_{b}\otimes{\vt e}_{4}+{\vt e}_{4}\otimes{\vt e}_{b}\right)
+t_{\rm{c4}}\left({\vt e}_{c}\otimes{\vt e}_{4}+{\vt e}_{4}\otimes{\vt e}_{c}\right)
+\tau{\vt e}_{4}\otimes{\vt e}_{4}
\end{array}
\end{equation}

The scalar quantities $T_{{\rm b}\perp}, T_{\rm b}, T_{\rm{bc}},
T_{\rm c}$, $t_{\rm{b4}}, t_{\rm{c4}}$, and $\tau$
correspond to the tensors ${\vt T}$, ${\vt t}$
and ${\vt \tau}$ from equations (\ref{elast}), (\ref{piez}) and
(\ref{diel}), respectively ($\otimes$ indicates dyadic multiplication).
They are obtained as:

\begin{equation}
\label{TTT-a}
T_{{\rm b}\perp} = {\cal C}_{66}\xi_{\rm b}^2+{\cal C}_{44}\xi_{\rm c}^2 ,
\end{equation}

\begin{equation}
\label{TTT-b}
T_{\rm b} = {\cal C}_{11}\xi_{\rm b}^2+{\cal C}_{44}\xi_{\rm c}^2 ,
\end{equation}

\begin{equation}
\label{TTT-c}
T_{\rm{bc}} = \left({\cal C}_{13}+{\cal C}_{44}\right)\xi_{\rm b}\xi_{\rm c} ,
\end{equation}

\begin{equation}
\label{TTT-d}
T_{\rm c} = {\cal C}_{44}\xi_{\rm b}^2+{\cal C}_{33}\xi_{\rm c}^2 ,
\end{equation}

\begin{equation}
\label{TTT-e}
t_{\rm{b4}} = \left(e_{31}+e_{15}\right)\xi_{\rm b}\xi_{\rm c} ,
\end{equation}

\begin{equation}
\label{TTT-f}
t_{\rm{c4}} = e_{15}\xi_{\rm b}^2+e_{33}\xi_{\rm c}^2 ,
\end{equation}

\begin{equation}
\label{TTT-g}
\tau = -\left(\eta_{11}\xi_{\rm b}^2+\eta_{33}\xi_{\rm c}^2\right)
\end{equation}

${\cal C}_{AB}=\left\{{\cal C}_{11}, {\cal C}_{44}, {\cal C}_{66}, {\cal C}_{13},
{\cal C}_{33}\right\}$ denote the elastic,
$e_{iA}=\left\{e_{15}, e_{31}, e_{33}\right\}$ the piezoelectric, and
$\eta_{ij}=\left\{\eta_{11}, \eta_{33}\right\}$ the dielectric moduli
of the hexagonal material. Subscripts $A,B$ represent
Voigt's notation whereas $i,j$ represent cartesian
subscripts, respectively. \\
${\vt {\cal T}}^{-1}\left({\vt \xi}\right)$ can be written as:

\begin{equation}
\label{Tinv}
{\vt {\cal T}}^{-1}\left({\vt \xi}\right) =
{\displaystyle \frac{{\vt \Lambda}\left({\vt \xi}\right)}
{f\left({\vt \xi}\right)}}
\end{equation}

Here $\Lambda_{ij}$ denotes the matrix of $3 \times 3$
subdeterminants (multiplied by the prefactor $(-1)^{i+j}$)
and $f\left({\vt \xi}\right)$ the determinant of ${\vt {\cal T}}$,
respectively. \\
We now can write for ${\vt \Lambda}$ by using
(\ref{userep}):

\begin{equation}
\label{Tinvrep}
\begin{array}{lll}
{\vt \Lambda}\left({\vt \xi}\right) &=&
\Lambda_{{\rm b}\perp}{\vt e}_{b\perp}\otimes{\vt e}_{b\perp}
+\Lambda_{\rm b}{\vt e}_{b}\otimes{\vt e}_{b}
+\Lambda_{\rm{bc}}\left({\vt e}_{b}\otimes{\vt e}_{c}+
{\vt e}_{c}\otimes{\vt e}_{b}\right)
+\Lambda_{\rm c}{\vt e}_{c}\otimes{\vt e}_{c} \nonumber \\ &&
+\Lambda_{\rm{b4}}\left({\vt e}_{b}\otimes{\vt e}_{4}+{\vt e}_{4}\otimes{\vt e}_{b}\right)
+\Lambda_{\rm{c4}}\left({\vt e}_{c}\otimes{\vt e}_{4}+{\vt e}_{4}\otimes{\vt e}_{c}\right)
+\Lambda_{\rm{4}}{\vt e}_{4}\otimes{\vt e}_{4}
\end{array}
\end{equation}

Here the scalar quantities are introduced:
\begin{equation}
\label{lam-a}
\Lambda_{{\rm b}\perp}=\tau\left(T_{\rm b}T_{\rm c}
-T_{\rm{bc}}^2\right)-\left(t_{\rm{c4}}^2T_{\rm b}-2t_{\rm{c4}}t_{\rm{b4}}T_{\rm{bc}}
+t_{\rm{b4}}^2T_{\rm c}\right) ,
\end{equation}

\begin{equation}
\label{lam-b}
\Lambda_{\rm b}=T_{{\rm b}\perp}\left(T_{\rm c}\tau-t_{\rm{c4}}^2\right) ,
\end{equation}

\begin{equation}
\label{lam-c}
\Lambda_{\rm{bc}}=-T_{{\rm b}\perp}\left(T_{\rm{bc}}\tau
-t_{\rm{b4}}t_{\rm{c4}}\right) ,
\end{equation}

\begin{equation}
\label{lam-d}
\Lambda_{\rm c}=T_{{\rm b}\perp}\left(T_{\rm b}\tau-t_{\rm{b4}}^2\right) ,
\end{equation}

\begin{equation}
\label{lam-e}
\Lambda_{\rm{b4}}=T_{{\rm b}\perp}\left(T_{\rm{bc}}t_{\rm{c4}}
-T_{\rm c}t_{\rm{b4}}\right) ,
\end{equation}

\begin{equation}
\label{lam-f}
\Lambda_{\rm{c4}} =-T_{{\rm b}\perp}\left(T_{\rm b}t_{\rm{c4}}
-T_{\rm{bc}}t_{\rm{b4}}\right) ,
\end{equation}

\begin{equation}
\label{lam-g}
\Lambda_{\rm{4}} =T_{{\rm b}\perp}\left(T_{\rm b}T_{\rm c}-T_{\rm{bc}}^2\right)
\end{equation}

The determinant $f\left({\vt \xi}\right)$ of ${\vt {\cal T}}$ then takes the
form

\begin{equation}
\label{deterrep}
f\left({\vt \xi}\right)=Det {\vt {\cal T}}\left({\vt \xi}\right)=T\left({\vt \xi}\right)_{{\rm b}\perp}
\Lambda_{{\rm b}\perp}\left({\vt \xi}\right)
\end{equation}

Putting equations (\ref{TTT-a})-(\ref{TTT-g}) into (\ref{deterrep}) shows that
only terms proportional $\xi_{\rm b}^{2n}\xi_{\rm c}^{8-2n}$
($n=0,1,2,3,4$) appear. This is a
unique property of the hexagonal medium. As we shall show this is
the central property which is needed to solve
our integration problem (\ref{repres}) explicitely.
Thus (\ref{lam-a})
is a polynomial of degree $3$ in $a$ when we put $a=\xi_{\rm b}^2/\xi_{\rm c}^2$.
We then can write

\begin{equation}
\label{poly}
\Lambda_{{\rm b}\perp}=P\left(a\right)\xi_{\rm c}^6
\end{equation}

where $P\left(a\right)$ is a polynomial of degree $3$ in $a$ and takes the
form:

\begin{equation}
\label{polynomy}
P\left(a\right)=Aa^3+Ba^2+Ca+D
\end{equation}

We obtain for the coefficients $A,B,C,D$:

\begin{equation}
\label{coeff-A}
\begin{array}{rcl}
A&=&-\eta_{11}{\cal C}_{11}{\cal C}_{44}-{\cal C}_{11}e_{15}^2
\end{array}
\end{equation}

\begin{equation}
\label{coeff-B}
\begin{array}{rcl}
B&=&-\eta_{33}{\cal C}_{11}{\cal C}_{44}
-\eta_{11}\left({\cal C}_{11}{\cal C}_{33}-2{\cal C}_{13}{\cal C}_{44}
-{\cal C}_{13}^2\right)-{\cal C}_{44}e_{15}^2-2{\cal C}_{11}e_{15}e_{33} \nonumber \\ &&
+2\left({\cal C}_{13}+{\cal C}_{44}\right)e_{15}\left(e_{31}+e_{15}\right)
-{\cal C}_{44}\left(e_{31}+e_{15}\right)^2
\end{array}
\end{equation}

\begin{equation}
\label{coeff-C}
\begin{array}{rcl}
C&=&-\eta_{33}\left({\cal C}_{11}{\cal C}_{33}
-2{\cal C}_{13}{\cal C}_{44}-{\cal C}_{13}^2\right)-\eta_{11}{\cal C}_{33}
{\cal C}_{44}-2e_{15}e_{33}{\cal C}_{44}-e_{33}^2{\cal C}_{11} \nonumber \\ &&
+2e_{33}\left(e_{31}+e_{15}\right)\left({\cal C}_{13}+{\cal C}_{44}\right)
-{\cal C}_{33}\left(e_{31}+e_{15}\right)^2
\end{array}
\end{equation}

\begin{equation}
\label{coeff-D}
\begin{array}{rcl}
D&=&-\eta_{33}{\cal C}_{33}{\cal C}_{44}-e_{33}^2{\cal C}_{44}
\end{array}
\end{equation}

Thus we can factorize the determinant $f$ according to:

\begin{equation}
\label{factor}
f\left({\vt \xi}\right)=\xi_{\rm c}^8{\cal C}_{66}A\left(a+A_1\right)
\left(a+A_2\right)\left(a+A_3\right)\left(a+A_4\right) ,
\end{equation}

with $T\left(a\right)_{{\rm b}\perp}={\cal C}_{66}\left(a+A_1\right)\xi_{\rm c}^2$,
($A_1={\cal C}_{44}/{\cal C}_{66}$) and

\begin{equation}
\label{polyfact}
P\left(a\right)=A\left(a+A_2\right)\left(a+A_3\right)\left(a+A_4\right)
\end{equation}

The $\Lambda's$ from equations (\ref{lam-a})-(\ref{lam-g}) yield
with $a=\xi_{\rm b}^2/\xi_{\rm c}^2$ and $\xi_c^2=1$:

\begin{equation}
\label{L-a}
\Lambda_{{\rm b}\perp}\left(a\right)=P\left(a\right)=Aa^3+Ba^2+Ca+D
\end{equation}

The numbers $A_2,A_3,A_4$ are the zeros of the equation:

\begin{equation}
\label{zeros}
Aa^3-Ba^2+Ca-D=0
\end{equation}
with above coefficients $A,B,C,D$ from equations
(\ref{coeff-A})-(\ref{coeff-D}). The $A_l$ are material quantities and
fully determined by the moduli ${\cal {\vt C}}, {\vt e}, {\vt \eta}$.
Furthermore, the subdeterminants (\ref{lam-b})-(\ref{lam-g}) yield:

\begin{equation}
\label{L-b}
\Lambda_{\rm b}\left(a\right)=-\left({\cal C}_{66}a+{\cal C}_{44}\right)
\left[\left(\eta_{11}a+\eta_{33}\right)\left({\cal C}_{44}a+{\cal C}_{33}\right)
+\left(e_{15}a+e_{33}\right)^2\right] ,
\end{equation}

\begin{equation}
\label{L-c}
\Lambda_{\rm{bc}}\left(a\right)=\sqrt{a}\left({\cal C}_{66}a+{\cal C}_{44}\right)
\left[\left(e_{31}+e_{15}\right)\left(e_{15}a+e_{33}\right)+
\left(\eta_{11}a+\eta_{33}\right)\left({\cal C}_{13}+{\cal C}_{44}\right)\right] ,
\end{equation}

\begin{equation}
\label{L-d}
\Lambda_{\rm c}\left(a\right)=-\left({\cal C}_{66}a+{\cal C}_{44}\right)
\left[\left(\eta_{11}a+\eta_{33}\right)\left({\cal C}_{11}a+{\cal C}_{44}\right)
+a\left(e_{31}+e_{15}\right)^2\right] ,
\end{equation}

\begin{equation}
\label{L-e}
\Lambda_{\rm{b4}}\left(a\right)=\sqrt{a}\left({\cal C}_{66}a+{\cal C}_{44}\right)
\left[\left({\cal C}_{13}+{\cal C}_{44}\right)\left(e_{15}a+e_{33}\right)
-\left({\cal C}_{44}a+{\cal C}_{33}\right)\left(e_{31}+e_{15}\right)\right] ,
\end{equation}

\begin{equation}
\label{L-f}
\Lambda_{\rm{c4}}\left(a\right)=-\left({\cal C}_{66}a+{\cal C}_{44}\right)
\left[\left({\cal C}_{11}a+{\cal C}_{44}\right)\left(e_{15}a+e_{33}\right)
-a\left({\cal C}_{13}+{\cal C}_{44}\right)\left(e_{31}+e_{15}\right)\right] ,
\end{equation}

\begin{equation}
\label{L-g}
\Lambda_{\rm{4}}\left(a\right)=\left({\cal C}_{66}a+{\cal C}_{44}\right)
\left[a^2{\cal C}_{11}{\cal C}_{44}
+a\left({\cal C}_{11}{\cal C}_{33}
-2{\cal C}_{13}{\cal C}_{44}-{\cal C}_{13}^2\right)
+{\cal C}_{33}{\cal C}_{44}\right]
\end{equation}

The dependence on $a$ is obtained by putting
$\xi_b=\sqrt{a}$ and $\xi_c=1$ in equations
(\ref{TTT-a})-(\ref{TTT-g}) and (\ref{lam-a})-(\ref{lam-g}), respectively.
\\ \\
To evaluate (\ref{repres}) we make use of the properties:

\begin{equation}
\label{homog6}
\Lambda_{ij}\left({\vt \xi}\lambda\right)=
\lambda^6\Lambda_{ij}\left({\vt \xi}\right)
\end{equation}

$\lambda$ denotes an arbitrary scalar number.
The components $\Lambda_{ij}$ are homogeneous functions of degree $6$.
Whereas the determinant $f$ is homogeneous of degree $8$:

\begin{equation}
\label{homog8}
f\left({\vt \xi}\lambda\right)=
\lambda^8f\left({\vt \xi}\right)
\end{equation}

Thus

\begin{equation}
\label{Tinho}
{\vt {\cal T}}^{-1}\left({\vt \xi}\right) =
{\displaystyle \lambda^2\frac{{\vt \Lambda}\left(\lambda{\vt \xi}\right)}
{f\left(\lambda{\vt \xi}\right)}}
\end{equation}

We now introduce the complex vector \cite{MiWuA}

\begin{equation}
\label{covec}
{\vt \gamma}\left(\alpha\right)=2e^{i\alpha}{\vt \xi}\left(\alpha\right)
\end{equation}
with ${\vt \xi}\left(\alpha\right)$ from equation (\ref{xi})
($i$ denotes the imaginary unit).
(\ref{covec}) can then be written as

\begin{equation}
\label{com}
{\vt \gamma}\left(s\right)={\vt h}^*s+{\vt h}
\end{equation}

Here we have introduced the complex variable

\begin{equation}
\label{comvar}
s=e^{2i\alpha}
\end{equation}
and the vector

\begin{equation}
\label{hdef}
{\vt h}={\vt e}_1+i{\vt e}_2
\end{equation}
with the basis vectors ${\vt e}_{1,2}$ from equations
(\ref{evec}). When putting $\lambda=2e^{i\alpha}$ equation (\ref{Tinho})
then can be written as:

\begin{equation}
\label{Tinhcom}
{\vt {\cal T}}^{-1}\left({\vt \xi}\right) =
{\displaystyle 4s\frac{{\vt \Lambda}\left({\vt \gamma}\left(s\right)\right)}
{f\left({\vt \gamma}\left(s\right)\right)}}
\end{equation}

Furthermore we observe

\begin{equation}
\label{des}
{\rm d}s=2is{\rm d}\alpha
\end{equation}

Then the integral (\ref{repres}) can be transformed into
a complex integral over the unit circle:

\begin{equation}
\label{egeu}
{\vt {\cal G}}\left({\vt r}\right) =
\frac{1}{8\pi^2r}
\oint_{|s|=1}\frac{4{\vt \Lambda}\left({\vt \gamma}
\left(s\right)\right)}
{f\left({\vt \gamma}\left(s\right)\right)}\frac{{\rm d}s}{i}
\end{equation}

Using the residue theorem we can rewrite integral (\ref{egeu})
as

\begin{equation}
\label{resform}
{\vt {\cal G}}\left({\vt r}\right) =
\frac{1}{8\pi^2r} 2\pi i\sum{{\rm Res}
\left(\frac{4{\vt \Lambda}\left({\vt \gamma}
\left(s\right)\right)}
{i f\left({\vt \gamma}\left(s\right)\right)}\right)}
\end{equation}

To evaluate (\ref{resform}) we have to find all zeros of
$f\left({\vt \gamma}\left(s\right)\right)$
which are located within the unit circle. To find these zeros we
observe that $f\left({\vt \gamma}\left(s\right)\right)$ is a polynomial
of degree $8$ in $s$. Thus there exist $8$ zeros $s_j$ of $f$.
We assume here that $f$ has no multiple zeros $s_j$. \\
Following \cite{MiWuA} we can conclude that there are $4$ pairs
of zeros $s_l, {\bar s}_l$ ($l=1,2,3,4$) having the property:

\begin{equation}
\label{basprop}
|s_l{\bar s}_l|=1
\end{equation}

There are only four zeros $s_l$ which are located within the unit circle.
Four zeros ${\bar s}_l$ lie outside the unit circle.
Thus the zeros can be written in the form \cite{MiWuA}:

\begin{equation}
\label{nustea}
s_l = {\rm e}^{2{\rm i}\Phi_l}{\rm e}^{-2\psi_l} ,
\end{equation}
\begin{equation}
\label{nusteb}
{\bar s}_l = {\rm e}^{2{\rm i}\Phi_l}{\rm e}^{+2\psi_l} ,
\end{equation}
where $\psi_l > 0$ ($l=1,2,3,4$). Here the degenerate case $|s_l|=|{\bar s}_l|$
is excluded by our assumption of no multible zeros.
Only the residues corresponding to the four zeros (\ref{nustea})
lying {\it within} the unit circle contribute to (\ref{resform}).
The residues at $s=s_l$ are obtained as:

\begin{equation}
\label{resi}
{\rm Res}
\left.\left(\frac{4{\vt \Lambda}\left({\vt \gamma}
\left(s\right)\right)}
{i f\left({\vt \gamma}\left(s\right)\right)}\right)\right|_{s=s_l}=
{\displaystyle \frac{4{\vt \Lambda}\left({\vt \gamma}
\left(s_l\right)\right)}
{\displaystyle {i \frac{{\rm d}f\left({\vt \gamma}\left(s\right)\right)}
{{\rm d}s}|_{s=s_l}}}}
\end{equation}

Finally we arrive at

\begin{equation}
\label{ressum}
{\vt {\cal G}}\left({\vt r}\right) =
\frac{1}{4\pi r} \sum_{l=1}^4{{\displaystyle
\frac{4{\vt \Lambda}\left({\vt \gamma}
\left(s_l\right)\right)}
{\displaystyle {\frac{{\rm d}f\left({\vt \gamma}\left(s\right)\right)}
{{\rm d}s}|_{s=s_l}}}}}
\end{equation}

The zeros $s_l$ of $f\left({\vt \gamma}\left(s\right)\right)$ within
the unit circle yield

\begin{equation}
\label{nuhex}
s_l = \frac{\sqrt{A_l\rho^2+z^2}-r}{\sqrt{A_l\rho^2+z^2}+r}
\end{equation}

To obtain (\ref{nuhex}) equation (\ref{factor}) together with
(\ref{com}) and (\ref{hdef}) is used.

\section{Explicit form of the Green's function}
Evaluating (\ref{ressum}) the Green's function takes the form:

\begin{equation}
\label{Green}
{\vt {\cal G}}\left({\vt r}\right)={\displaystyle
\frac{1}{4\pi A{\cal C}_{66}}\sum_{l=1}^{4}
{\frac{{\vt \Lambda}\left({\vt \xi}^{\left(l\right)}\right)}
{\sqrt{A_l\rho^2+z^2}\prod_{j=1,(j\neq l)}^{4}{\left(A_j-A_l\right)}}}}
\end{equation}

${\vt \xi}^{\left(l\right)}$ ($l=1,2,3,4$) are the solutions of the equations

\begin{equation}
\label{fnull}
f\left({\vt \xi}\right)=0
\end{equation}

or

\begin{equation}
\label{zuscon}
\xi_1^2+\xi_2^2+A_l\xi_3^2=0
\end{equation}

and

\begin{equation}
\label{rnull}
{\vt \xi}{\vt r}=\xi_1x+\xi_2y+\xi_3z=0
\end{equation}
with $\xi_3^{\left(l\right)}=1$. Thus equations (\ref{fnull}) and (\ref{rnull})
have the solutions ($l=1,2,3,4$):

\begin{equation}
\label{sol1}
\xi_1^{\left(l\right)}=\frac{1}{\rho^2}\left[-zx+iy\sqrt{A_l\rho^2+z^2}\right]
\end{equation}
\begin{equation}
\label{sol2}
\xi_2^{\left(l\right)}=\frac{1}{\rho^2}\left[-zy-ix\sqrt{A_l\rho^2+z^2}\right]
\end{equation}

($\rho=\sqrt{x^2+y^2}$).
This formulation is in complete analogy to the result obtained by
Kr\"oner for the Green's function tensor of the hexagonal elastic medium
\cite{Kroener4} by using Fredholm's method
\cite{Fredholm}. \\ \\
The Green's function (\ref{Green}) then assumes the form:

\begin{equation}
\label{Greenform}
\begin{array}{lll}
{\vt {\cal G}}\left({\vt r}\right) &=&
{\cal G}_{\phi \phi}\left(\rho,z\right){\vt e}_{\phi}\otimes{\vt e}_{\phi}
+{\cal G}_{\rho \rho}\left(\rho,z\right){\vt e}_{\rho}\otimes{\vt e}_{\rho}
+{\cal G}_{\rm{\rho z}}\left(\rho,z\right)\left({\vt e}_{\rho}\otimes{\vt e}_{z}
+{\vt e}_{z}\otimes{\vt e}_{\rho}\right) \nonumber \\ &&
+{\cal G}_{z z}\left(\rho,z\right){\vt e}_{z}\otimes{\vt e}_{z}
+{\cal G}_{\rm{\rho 4}}\left(\rho,z\right)\left({\vt e}_{\rho}\otimes{\vt e}_{4}+
{\vt e}_{4}\otimes{\vt e}_{\rho}\right)  \nonumber \\ &&
+{\cal G}_{z 4}\left(\rho,z\right)\left({\vt e}_{z}\otimes{\vt e}_{4}
+{\vt e}_{4}\otimes{\vt e}_{z}\right)
+{\cal G}_{4 4}\left(\rho,z\right){\vt e}_{4}\otimes{\vt e}_{4}
\end{array}
\end{equation}

Here we have introduced the following basis:

\begin{equation}
\label{hexbas}
\begin{array}{lcrclcrclcrclcr}
  {\vt e}_{\rho} & = & \frac{\displaystyle 1}{\displaystyle \rho}
  \left( \begin{array}{c}
       x \\ y \\ 0\\0
\end{array}    \right) &,&
  {\vt e}_{\phi} & = &\frac{\displaystyle 1}{\displaystyle \rho}
  \left( \begin{array}{c}
     -y \\ x \\ 0 \\ 0
\end{array}    \right) &,&
  {\vt e}_{z} & = &
  \left( \begin{array}{c}
             0 \\ 0 \\ 1 \\0
\end{array}    \right) &,&
  {\vt e}_{4} & = &
  \left( \begin{array}{c}
             0 \\ 0 \\ 0 \\1
\end{array}    \right) \\
\end{array}
\end{equation}

To obtain (\ref{Greenform}) we have used (\ref{Tinvrep}) together with
(\ref{sol1}) and (\ref{sol2}).
Equation (\ref{Green}) then yields the ${\cal G}'s$ introduced in
(\ref{Greenform}) as:

\begin{equation}
\label{Gfi}
{\cal G}_{\phi \phi}\left(\rho,z\right)={\displaystyle
\frac{1}{4\pi A{\cal C}_{66}}\sum_{l=1}^{4}
{\frac{
\rho^2\Lambda_{\rm b}\left(a=-A_l\right)+z^2\Gamma_{\rm b}\left(a=-A_l\right)}
{\rho^2\sqrt{A_l\rho^2+z^2}\prod_{j=1,(j\neq l)}^{4}{\left(A_j-A_l\right)}}}}
\end{equation}

\begin{equation}
\label{Gro}
{\cal G}_{\rho \rho}\left(\rho,z\right)={\displaystyle
\frac{1}{4\pi A{\cal C}_{66}}\sum_{l=1}^{4}
{\frac{
\rho^2\Lambda_{{\rm b}\perp}\left(a=-A_l\right)-z^2\Gamma_{\rm b}\left(a=-A_l\right)}
{\rho^2\sqrt{A_l\rho^2+z^2}\prod_{j=1,(j\neq l)}^{4}{\left(A_j-A_l\right)}}}}
\end{equation}

\begin{equation}
\label{Groz}
{\cal G}_{\rho z}\left(\rho,z\right)={\displaystyle
\frac{1}{4\pi A{\cal C}_{66}}\sum_{l=1}^{4}
{\frac{\left(-z\right)\Gamma_{\rm{bc}}\left(a=-A_l\right)}
{\rho\sqrt{A_l\rho^2+z^2}\prod_{j=1,(j\neq l)}^{4}{\left(A_j-A_l\right)}}}}
\end{equation}

\begin{equation}
\label{Gro4}
{\cal G}_{\rho 4}\left(\rho,z\right)={\displaystyle
\frac{1}{4\pi A{\cal C}_{66}}\sum_{l=1}^{4}
{\frac{\left(-z\right)\Gamma_{\rm{b4}}\left(a=-A_l\right)}
{\rho\sqrt{A_l\rho^2+z^2}\prod_{j=1,(j\neq l)}^{4}{\left(A_j-A_l\right)}}}}
\end{equation}

\begin{equation}
\label{Grzz}
{\cal G}_{z z}\left(\rho,z\right)={\displaystyle
\frac{1}{4\pi A{\cal C}_{66}}\sum_{l=1}^{4}
{\frac{\Lambda_{\rm{ c}}\left(a=-A_l\right)}
{\sqrt{A_l\rho^2+z^2}\prod_{j=1,(j\neq l)}^{4}{\left(A_j-A_l\right)}}}}
\end{equation}

\begin{equation}
\label{Grz4}
{\cal G}_{z 4}\left(\rho,z\right)={\displaystyle
\frac{1}{4\pi A{\cal C}_{66}}\sum_{l=1}^{4}
{\frac{\Lambda_{\rm{c4}}\left(a=-A_l\right)}
{\sqrt{A_l\rho^2+z^2}\prod_{j=1,(j\neq l)}^{4}{\left(A_j-A_l\right)}}}}
\end{equation}

\begin{equation}
\label{Gr44}
{\cal G}_{4 4}\left(\rho,z\right)={\displaystyle
\frac{1}{4\pi A{\cal C}_{66}}\sum_{l=1}^{4}
{\frac{\Lambda_{\rm{ 4}}\left(a=-A_l\right)}
{\sqrt{A_l\rho^2+z^2}\prod_{j=1,(j\neq l)}^{4}{\left(A_j-A_l\right)}}}}
\end{equation}

Here we have introduced the quantities

\begin{equation}
\label{gabdef}
\Lambda_{{\rm b}\perp}\left(a\right)-\Lambda_{\rm b}\left(a\right)=a\Gamma_{\rm b}\left(a\right)
\end{equation}

to arrive at

\begin{equation}
\label{gab}
\begin{array}{lcl}
\Gamma_{\rm b}\left(a\right)&=&\left({\cal C}_{11}-{\cal C}_{66}\right)
\left(T_{\rm c}\left(a\right)\tau\left(a\right)
-t_{\rm{c4}}^2\left(a\right)\right) \nonumber \\ &&
-\left({\cal C}_{13}+{\cal C}_{44}\right)^2\tau\left(a\right)
+2\left({\cal C}_{13}+{\cal C}_{44}\right)
\left(e_{31}+e_{15}\right)t_{\rm{c4}}\left(a\right)
\nonumber \\ &&
+\left(e_{31}+e_{15}\right)^2T_{\rm c}\left(a\right)
\end{array}
\end{equation}

and

\begin{equation}
\label{gabc}
\Lambda_{\rm{bc}}\left(a\right)=\sqrt{a}\Gamma_{\rm{bc}}\left(a\right)
\end{equation}

where (compare eq. (\ref{L-c}))

\begin{equation}
\label{L-c-g}
\Gamma_{\rm{bc}}\left(a\right)=\left({\cal C}_{66}a+{\cal C}_{44}\right)
\left[\left(e_{31}+e_{15}\right)\left(e_{15}a+e_{33}\right)+
\left(\eta_{11}a+\eta_{33}\right)\left({\cal C}_{13}+{\cal C}_{44}\right)\right] ,
\end{equation}

and

\begin{equation}
\label{gab4def}
\Lambda_{\rm{b4}}\left(a\right)=\sqrt{a}\Gamma_{\rm{b4}}\left(a\right)
\end{equation}

where (compare eq. (\ref{L-e}))

\begin{equation}
\label{L-e-g}
\Gamma_{\rm{b4}}\left(a\right)=\left({\cal C}_{66}a+{\cal C}_{44}\right)
\left[\left({\cal C}_{13}+{\cal C}_{44}\right)\left(e_{15}a+e_{33}\right)
-\left({\cal C}_{44}a+{\cal C}_{33}\right)\left(e_{31}+e_{15}\right)\right]
\end{equation}

Equation (\ref{Greenform}) together with (\ref{Gfi})-(\ref{Gr44})
represents the electroelastic Green's function explicitely in compact form.
Especially for the calculation of the electroelastic Eshelby tensor
the use of representation (\ref{Greenform}) may be convenient. \\

The cartesian representation of the the electroelastic Green's function
becomes:

\begin{equation}
\label{Greentensor}
\begin{array}{ll}
{\vt {\cal G}}\left({\vt r}\right) = {\displaystyle
{\sum_{l=1}^{4}\frac{1}{\sqrt{A_l\rho^2+z^2}}}}
\times&\left( \begin{array}{llll}
g_{11}^{(l)}
&g_{12}^{(l)}&g_{13}^{(l)}&g_{14}^{(l)} \\g_{12}^{(l)}
&g_{22}^{(l)}
&g_{23}^{(l)}&g_{24}^{(l)} \\
g_{13}^{(l)}&
g_{23}^{(l)}&g_{33}^{(l)}&g_{34}^{(l)} \\
g_{14}^{(l)}&g_{24}^{(l)}&g_{34}^{(l)}&g_{44}^{(l)} \\
\end{array} \right) \\
\end{array}
\end{equation}

Here we have used the abbreviations:

\begin{equation}
\label{g11}
g_{11}^{(l)}={\displaystyle \frac{1}{{\cal E}_l}}\left[{\displaystyle {-\Gamma_{\rm b}\left(-A_l\right)\frac{x^2z^2-y^2\left(A_l\rho^2+z^2\right)}
{\rho^4}+\Lambda_{{\rm b}\perp}\left(-A_l\right)}}\right]
\end{equation}

\begin{equation}
\label{g22}
g_{22}^{(l)}={\displaystyle \frac{1}{{\cal E}_l}}\left[{\displaystyle
{-\Gamma_{\rm b}\left(-A_l\right)\frac{y^2z^2-x^2\left(A_l\rho^2+z^2\right)}
{\rho^4}+\Lambda_{{\rm b}\perp}\left(-A_l\right)}}\right]
\end{equation}

\begin{equation}
\label{g12}
g_{12}^{(l)}={\displaystyle \frac{1}{{\cal E}_l}}\left[{\displaystyle {-\Gamma_{\rm b}\left(-A_l\right)
\frac{xy\left(A_l\rho^2+2z^2\right)}{\rho^4}}}\right]
\end{equation}

\begin{equation}
\label{g13}
g_{13}^{(l)}={\displaystyle \frac{1}{{\cal E}_l}}\left[{\displaystyle -\Gamma_{\rm{bc}}
\left(-A_l\right)\frac{xz}{\rho^2}}\right]
\end{equation}

\begin{equation}
\label{g23}
g_{23}^{(l)}={\displaystyle \frac{1}{{\cal E}_l}}\left[{\displaystyle -\Gamma_{\rm{bc}}
\left(-A_l\right)\frac{yz}{\rho^2}}\right]
\end{equation}

\begin{equation}
\label{g33}
g_{33}^{(l)}={\displaystyle \frac{1}{{\cal E}_l}}{\displaystyle \Lambda_{\rm c}
\left(-A_l\right)}
\end{equation}

\begin{equation}
\label{g14}
g_{14}^{(l)}={\displaystyle \frac{1}{{\cal E}_l}}\left[{\displaystyle -\Gamma_{\rm{b4}}
\left(-A_l\right)\frac{xz}{\rho^2}}\right]
\end{equation}

\begin{equation}
\label{g24}
g_{24}^{(l)}={\displaystyle \frac{1}{{\cal E}_l}}\left[{\displaystyle -\Gamma_{\rm{b4}}
\left(-A_l\right)\frac{yz}{\rho^2}}\right]
\end{equation}

\begin{equation}
\label{g34}
g_{34}^{(l)}={\displaystyle \frac{1}{{\cal E}_l}}{\displaystyle \Lambda_{\rm{c4}}\left(-A_l\right)}
\end{equation}

\begin{equation}
\label{g44}
g_{44}^{(l)}={\displaystyle \frac{1}{{\cal E}_l}}{\displaystyle \Lambda_{\rm 4}\left(-A_l\right)}
\end{equation}

and

\begin{equation}
\label{epsel}
{\cal E}_l = 4\pi{\cal C}_{66}A\prod_{j=1,(j\neq l)}^{4}{\left(A_j-A_l\right)}
\end{equation}

As we shall show in the appendix the Green's
function (\ref{Greentensor}) yields in the case of vanishing
piezoelectric coupling $e_{ijk}=0$ the well known results:
For the elastic part ${\cal G}_{ij}$
(i,j=1,2,3) Kr\"oner's elastic Green's tensor
of the hexagonal medium \cite{Kroener4} and the dielectric part
${\cal G}_{44}$ then represents the solution of Poisson equation
of a unit point charge in a hexagonal dielectric medium,
whereas the ${\cal G}_{j4}$ ($j=1,2,3$) are vanishing.

\section{Appendix}
Here we consider the case of vanishing piezoelectric coupling
($e_{ijk}=0$).
As a consequence the terms (eqs. (\ref{TTT-e}), (\ref{TTT-f}))

\begin{equation}
\label{coseq}
t_{\rm{b4}}=t_{\rm{c4}}=0
\end{equation}
are vanishing. Thus the determinant $f$ (eq. (\ref{deterrep})) simplifies according to

\begin{equation}
\label{detokop}
f\left(a\right)=Det {\vt {\cal T}}\left(a\right)=
\tau\left(a\right)T_{{\rm b}\perp}\left(a\right)
\left(T_{\rm b}\left(a\right)T_{\rm c}\left(a\right)-
T_{\rm{bc}}^2\left(a\right)\right)
\end{equation}
and can be written as

\begin{equation}
\label{defa}
f\left(a\right)=-\eta_{11}{\cal C}_{11}{\cal C}_{44}{\cal C}_{66}
\left(a+a_1\right)\left(a+a_2\right)\left(a+a_3\right)\left(a+a_4\right)
\end{equation}
where $a_l=A_l$ denote the zeros of $f\left(-a\right)$ in the decoupled case.

\begin{equation}
\label{a1}
a_1=\frac{{\cal C}_{44}}{{\cal C}_{66}}
\end{equation}
represents the zero of $T_{{\rm b}\perp}\left(-a\right)$.
The numbers $a_{2,3}$ are the zeros of the quadratic equation

\begin{equation}
\label{quadr}
T_{\rm b}\left(-a\right)T_{\rm c}\left(-a\right)-
T_{\rm{bc}}^2\left(-a\right)=
{\cal C}_{11}{\cal C}_{44}a^2+\left({\cal C}_{13}^2+2{\cal C}_{13}{\cal C}_{44}-{\cal C}_{11}{\cal C}_{33}\right)a+{\cal C}_{33}{\cal C}_{44}
=0
\end{equation}

The zeros $a_l$ for $l=1,2,3$ (elastic part) are those introduced in \cite{Kroener4}.
$a_4$ is given by

\begin{equation}
\label{a4}
a_4=\frac{\eta_{33}}{\eta_{11}}
\end{equation}
and represents the zero of $\tau\left(-a\right)$.
Furthermore we find when using (\ref{L-a})-(\ref{L-g}) together with
(\ref{gabdef})-(\ref{L-e-g})

\begin{equation}
\label{lo-a}
\Lambda_{{\rm b}\perp}\left(a=-a_l\right)=
-\eta_{11}{\cal C}_{11}{\cal C}_{44}\left(a_2-a_l\right)
\left(a_3-a_l\right)\left(a_4-a_l\right)
\end{equation}

\begin{equation}
\label{lo-b}
\Lambda_{{\rm b}}\left(a=-a_l\right)=
-\eta_{11}{\cal C}_{66}\left({\cal C}_{33}-a_l{\cal C}_{44}\right)
\left(a_1-a_l\right)
\left(a_4-a_l\right)
\end{equation}

\begin{equation}
\label{Gam}
\Gamma_{\rm b}\left(a=-a_l\right)=\eta_{11}\left(a_4-a_l\right)
\left[\left({\cal C}_{66}-{\cal C}_{11}\right)\left({\cal C}_{33}
-a_l{\cal C}_{44}\right)+\left({\cal C}_{13}+{\cal C}_{44}\right)^2\right]
\end{equation}

\begin{equation}
\label{lga-c}
\Gamma_{\rm{bc}}\left(a=-a_l\right)=
\eta_{11}{\cal C}_{66}\left({\cal C}_{13}+{\cal C}_{44}\right)
\left(a_1-a_l\right)\left(a_4-a_l\right)
\end{equation}

\begin{equation}
\label{lo-d}
\Lambda_{\rm c}\left(a=-a_l\right)=
-\eta_{11}{\cal C}_{66}\left({\cal C}_{44}-a_l{\cal C}_{11}\right)
\left(a_1-a_l\right)\left(a_4-a_l\right)
\end{equation}

and

\begin{equation}
\label{galo}
\Gamma_{\rm{b4}}\left(a=-a_l\right)=\Lambda_{\rm{c4}}\left(a=-a_l\right)=0
\end{equation}

to obtain ${\cal G}_{j4}=0$ ($j=1,2,3$) and

\begin{equation}
\label{LA-4}
\Lambda_{\rm 4}\left(a=-a_l\right)={\cal C}_{11}{\cal C}_{44}{\cal C}_{66}
\left(a_1-a_l\right)\left(a_2-a_l\right)\left(a_3-a_l\right)
\end{equation}

and for $l=1,2,3$:

\begin{equation}
\label{eps}
{\cal E}_l = -\eta_{11}\left(a_4-a_l\right){\rm E}_l
\end{equation}

with

\begin{equation}
\label{Elk}
{\rm E}_l=4\pi{\cal C}_{11}{\cal C}_{44}{\cal C}_{66}\prod_{j=1,(j\neq l)}^{3}{\left(a_j-a_l\right)}
\end{equation}

The terms (\ref{Elk}) coincide with Kr\"oner's (corrected) $"{\rm E}_l"$
(the terms ${\rm E}_l$ defined in \cite{Kroener4} have to
be corrected by a prefactor $-{\cal C}_{11}{\cal C}_{44}{\cal C}_{66}$
to obtain the correct result for the elastic Green's tensor \cite{Yoo}).\\
For $l=4$ we obtain:

\begin{equation}
\label{elep}
{\cal E}_4 = -4\pi\eta_{11}{\cal C}_{11}{\cal C}_{44}{\cal C}_{66}
\left(a_1-a_4\right)\left(a_2-a_4\right)\left(a_3-a_4\right)=
-4\pi\eta_{11}\Lambda_{\rm 4}\left(-a_4\right)
\end{equation}

We observe from equation (\ref{lo-a})-(\ref{lo-d}) the properties:

\begin{equation}
\label{agitprop1}
\frac{\Lambda_{{\rm b}\perp}\left(a=-a_4\right)}{{\cal E}_4}=
\frac{\Lambda_{\rm b}\left(a=-a_4\right)}{{\cal E}_4}=
\frac{\Gamma_{\rm b}\left(a=-a_4\right)}{{\cal E}_4}
=\frac{\Gamma_{\rm{bc}}\left(a=-a_4\right)}{{\cal E}_4}=
\frac{\Lambda_{\rm c}\left(a=-a_4\right)}{{\cal E}_4}=0
\end{equation}

(${\cal E}_4 \neq 0$, eq. (\ref{elep})). Thus the term $l=4$
in sum (\ref{Greentensor}) does not contribute to the elastic
components ${\cal G}_{ij}$ ($i,j=1,2,3$).
Using equations (\ref{LA-4}) and (\ref{eps}) together with (\ref{Elk})
we obtain for $l=1,2,3$:

\begin{equation}
\label{agitprop2}
\frac{\Lambda_{\rm 4}\left(a=-a_l\right)}{{\cal E}_l}=0
\end{equation}

Thus there are no contributions
to sum (\ref{Greentensor}) for $l=1,2,3$ to the dielectric part
${\cal G}_{44}$. For $l=4$ we find from equations (\ref{LA-4})
and (\ref{elep}):

\begin{equation}
\label{agitprop3}
\frac{\Lambda_{\rm 4}\left(a=-a_4\right)}{{\cal E}_4}=\frac{-1}{4\pi\eta_{11}}
\end{equation}

(\ref{agitprop3}) is independent on the elastic moduli ${\cal C}_{ijkl}$ which is a consequence
of $e_{ijk}=0$. Thus the dielectric part ${\cal G}_{44}$ of the Green's function
(\ref{Greentensor}) yields

\begin{equation}
\label{G44}
{\cal G}_{44}\left({\vt r}\right)=
\frac{\Lambda_{\rm 4}\left(a=-a_4\right)}{{\cal E}_4}
\frac{1}{\sqrt{a_4\rho^2+z^2}} = \frac{-1}{4\pi\eta_{11}\sqrt{a_4\rho^2+z^2}}
\end{equation}
where $a_4=\eta_{33}/\eta_{11}$. Indeed it is easily checked that (\ref{G44})
is the solution of the Poisson equation of a unit point charge
(compare (\ref{diel}) and (\ref{Greendef}))

\begin{equation}
\label{Poisson}
\tau\left(\nabla\right){\cal G}_{44}\left({\vt r}\right)+
\delta^3\left({\vt r}\right)=0
\end {equation}
where (compare equation (\ref{diel}))

\begin{equation}
\label{deta}
\tau\left(\nabla\right)=-{\displaystyle \left[\eta_{11}
\left(\frac{\partial^2}{\partial x^2}+\frac{\partial^2}{\partial y^2}\right)
+\eta_{33}\frac{\partial^2}{\partial z^2}\right]}
\end{equation}

Let us consider now the cases $l=1,2,3$:\\
From equations (\ref{lo-a})-(\ref{lo-d}) together with
(\ref{eps}) we obtain:

\begin{equation}
\label{elaprop2}
-\frac{\Gamma_{\rm b}\left(a=-a_l\right)}{{\cal E}_l}={\rm {\cal A}}_l=
\frac{\left({\cal C}_{66}-{\cal C}_{11}\right)
\left({\cal C}_{33}-a_l{\cal C}_{44}\right)+
\left({\cal C}_{13}+{\cal C}_{44}\right)^2}{{\rm E}_l}
\end{equation}

These terms coincide with Kr\"oner's $"{\rm {\cal A}}_l"$ from \cite{Kroener4}.
Furthermore we obtain:

\begin{equation}
\label{elaprop1}
\frac{\Lambda_{{\rm b}\perp}\left(a=-a_l\right)}{{\cal E}_l}={\rm B}_l=
\frac{{\cal C}_{11}{\cal C}_{44}a_l^2+\left({\cal C}_{13}^2+
2{\cal C}_{13}{\cal C}_{44}-{\cal C}_{11}{\cal C}_{33}\right)a_l
+{\cal C}_{33}{\cal C}_{44}}{{\rm E}_l}
\end{equation}

These terms coincide with Kr\"oner's $"{\rm B}_l"$ from \cite{Kroener4}.
\\
Furthermore we obtain

\begin{equation}
\label{elaprop3}
-\frac{\Gamma_{\rm{bc}}\left(a=-a_l\right)}{{\cal E}_l}={\rm C}_l=
\frac{\left({\cal C}_{44}-a_l{\cal C}_{66}\right)
\left({\cal C}_{13}+{\cal C}_{44}\right)}{{\rm E}_l}
\end{equation}

These terms coincide with Kr\"oner's $"{\rm C}_l"$ from \cite{Kroener4}.\\
Finally we obtain

\begin{equation}
\label{elaprop4}
\frac{\Lambda_{\rm c}\left(a=-a_l\right)}{{\cal E}_l}={\rm D}_l=\frac{
\left({\cal C}_{44}-a_l{\cal C}_{66}\right)
\left({\cal C}_{44}-a_l{\cal C}_{11}\right)}{{\rm E}_l}
\end{equation}

These terms concide with Kr\"oner's $"{\rm D}_l"$ from \cite{Kroener4}.\\
Because of the property (\ref{agitprop1}) the elastic part of the
Green's function ${\cal G}_{ij}=G_{ij}$ ($i,j=1,2,3$) yields by putting
(\ref{elaprop2})-(\ref{elaprop4}) into (\ref{Greentensor}) and by using
the abbreviations (\ref{g11})-(\ref{g33}) Kr\"oner's
result \cite{Kroener4}:

\begin{equation}
\label{GreenscherTensor}
\begin{array}{ll}
{\rm {\bf G}}\left({\vt r}\right) = {\displaystyle
{\sum_{l=1}^{3}\frac{1}{\sqrt{a_l\rho^2+z^2}}}}\times &\\ \\
 \left( \begin{array}{lll}
{\displaystyle {{\rm {\cal A}}_l\frac{x^2z^2-y^2\left(a_l\rho^2+z^2\right)}
{\rho^4}+{\rm B}_l}};
&{\displaystyle {{\rm {\cal A}}_l\frac{xy\left(a_l\rho^2+2z^2\right)}{\rho^4}}};
&{\displaystyle {\rm C}_l\frac{xz}{\rho^2}} \\{\displaystyle
{{\rm {\cal A}}_l\frac{xy\left(a_l\rho^2+2z^2\right)}{\rho^4}}};
&{\displaystyle
{{\rm {\cal A}}_l\frac{y^2z^2-x^2\left(a_l\rho^2+z^2\right)}{\rho^4}+{\rm B}_l}};
&{\displaystyle {{\rm C}_l\frac{yz}{\rho^2}}} \\
{\displaystyle {{\rm C}_l\frac{xz}{\rho^2}}}; &
{\displaystyle {{\rm C}_l\frac{yz}{\rho^2}}}; &{\displaystyle {{\rm D}_l}} \\
\end{array} \right) \\
\end{array}
\end{equation}
\newline

Here we used Kr\"oner's notation. Above constants ${\rm {\cal A}}_l,
{\rm B}_l, {\rm C}_l, {\rm D}_l$ are defined in equations
(\ref{elaprop2})-(\ref{elaprop4}), respectively.

\section{Conclusion}
The electroelastic 4$\times$4 Green's function of a piezoelectric
hexagonal medium which is infinitely extended has been calculated
explicitely by using residue theory. The obtained expression
(\ref{Greentensor}) is highly convenient for a treatment of the electroelastic
Eshelby tensor \cite{KuhnMichel}.\\
An important future application may be the following: The obtained
Green's function will be an important quantity even for a
treatment of nonlinear, e.g. hysteretic material behavior
in piezoelectric ceramics. To model these effects has become of great interest
since the increasing technological importance of
piezoelectric materials, in particular as electromechanical
actuators and sensors \cite{NanClarke,Kamlah,MichelKreher}.
Key contributions for a
treatment of such nonlinear effects using
the technique of Green's functions had been presented by
Wunderlin and Haken \cite{WunderHaken,HakenWunder}.

\begin{center}
{\bf \it{Acknowledgements}}
\end{center}
The author is grateful to Dr. W. Kreher, Professors V. M. Levin
and E. Kr\"oner for
helpful discussions
and their steady interest in this work and to R. Kuhn for checking some
expressions with MATHEMATICA.

\end{document}